\documentclass[a4paper,11pt]{article}
\usepackage{pos}
\usepackage{booktabs}

\newcommand{\ttbar}{\ensuremath{\mathrm{t} \bar{\mathrm{t}}}}
\newcommand{\ttg}{\ensuremath{\ttbar \gamma}}
\newcommand{\twg}{\ensuremath{\mathrm{tW} \gamma}}
\renewcommand{\ttz}{\ensuremath{\ttbar \mathrm{Z}}}
\newcommand{\tzq}{\ensuremath{\mathrm{t} \mathrm{Z} \mathrm{q}}}
\newcommand{\tqg}{\ensuremath{\mathrm{t} \mathrm{q} \gamma}}
\newcommand{\pt}{\ensuremath{p_\mathrm{T}}}

\newcommand{\delR}{\ensuremath{\Delta{\mathrm{R}}}}

\title{Measurement of top-quark electroweak couplings in associated top quark production with vector bosons at the ATLAS and CMS experiments}
\ShortTitle{Measurements of \ttg,\ \ttz,\ \tzq, and \tqg\ at the ATLAS and CMS}

\author*[1]{David Walter}
\note{On behalf of the ATLAS and CMS Collaborations}

\affiliation[]{Deutsches Electronen-Synchrotron (DESY),\\
  Notkestraße 85, Hamburg, Germany}

\emailAdd{david.walter@cern.ch}

\abstract{
Recent analyses of top quark production in association with vector bosons are summarized, representing the most precise inclusive and differential cross section measurements of these processes to date. Proton-proton collision data at a center-of-mass energy of $\sqrt{s}=13$\,TeV, recorded by the ATLAS and CMS detectors at the LHC are analyzed, corresponding to an integrated luminosity of up to 139\,fb$^{-1}$ for each experiment. Comparisons with theory calculations are performed and overall good agreement with standard model predictions are obtained.}

\FullConference{%
  *** The European Physical Society Conference on High Energy Physics (EPS-HEP2021), ***\\
  *** 26-30 July 2021 ***\\
  *** Online conference, jointly organized by Universität Hamburg and the research center DESY ***
}

\begin{document}
\maketitle

\section{Introduction}
During LHC Run-2 from 2015 to 2018, proton-proton collisions were performed at the highest collider energy $\sqrt{s}=13$\,TeV to date. 
In the process, the ATLAS~\cite{ATLAS} and CMS~\cite{CMS} experiments collected an unprecedented amount of data corresponding to an integrated luminosity of 139\,fb$^{-1}$ and 138\,fb$^{-1}$, respectively.
For the first time, detailed studies of the production of top quarks in association with photons (\tqg\ and \ttg) and Z bosons (\ttz\ and \tzq) became possible. The study of these processes is of particular interest because the electroweak (EW) couplings of the top quark to the photon (top-$\gamma$) or Z boson (top-Z) are sensitive at tree-level to beyond-the-standard-model (BSM) physics like flavor-changing neutral currents (FCNC) or vector-like quarks. Deviations from the standard model (SM) prediction can be parameterized as anomalous couplings or in effective field theory (EFT). Furthermore, a good understanding of these processes as backgrounds is needed for measurements of four top quarks or processes with Higgs bosons.
Presented are the most precise inclusive cross section measurements to date: first differential cross section measurements and new limits on beyond-the-standard-model effects.

\section{Top quark pair production in association with a photon}
Inclusive and differential cross section measurements of \ttg\ production have been performed by both ATLAS and CMS experiments, using full Run-2 data. 

\begin{figure}[!hbtp]
\centering
    \includegraphics[width=0.54\textwidth]{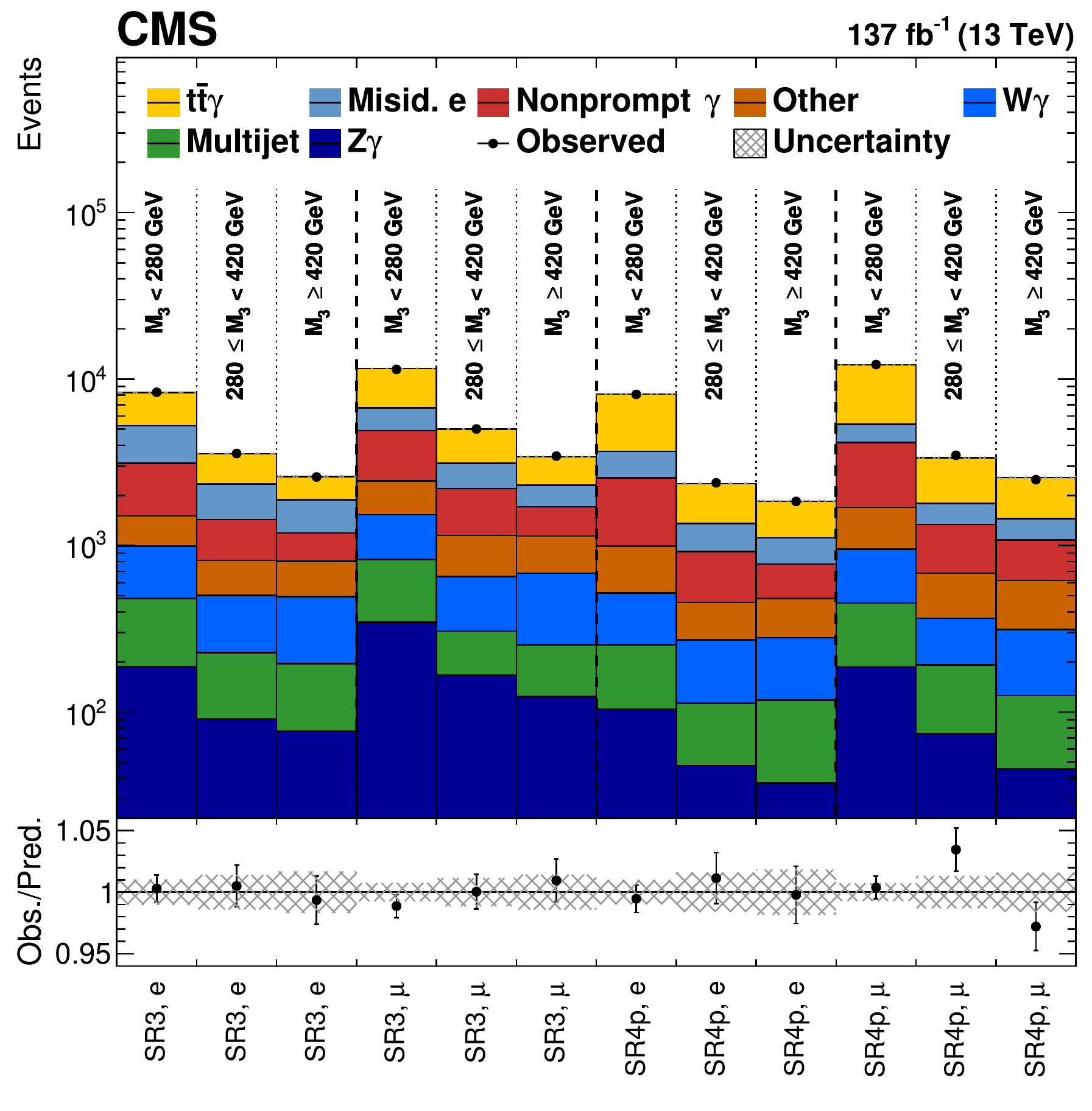}
    \includegraphics[width=0.42\textwidth]{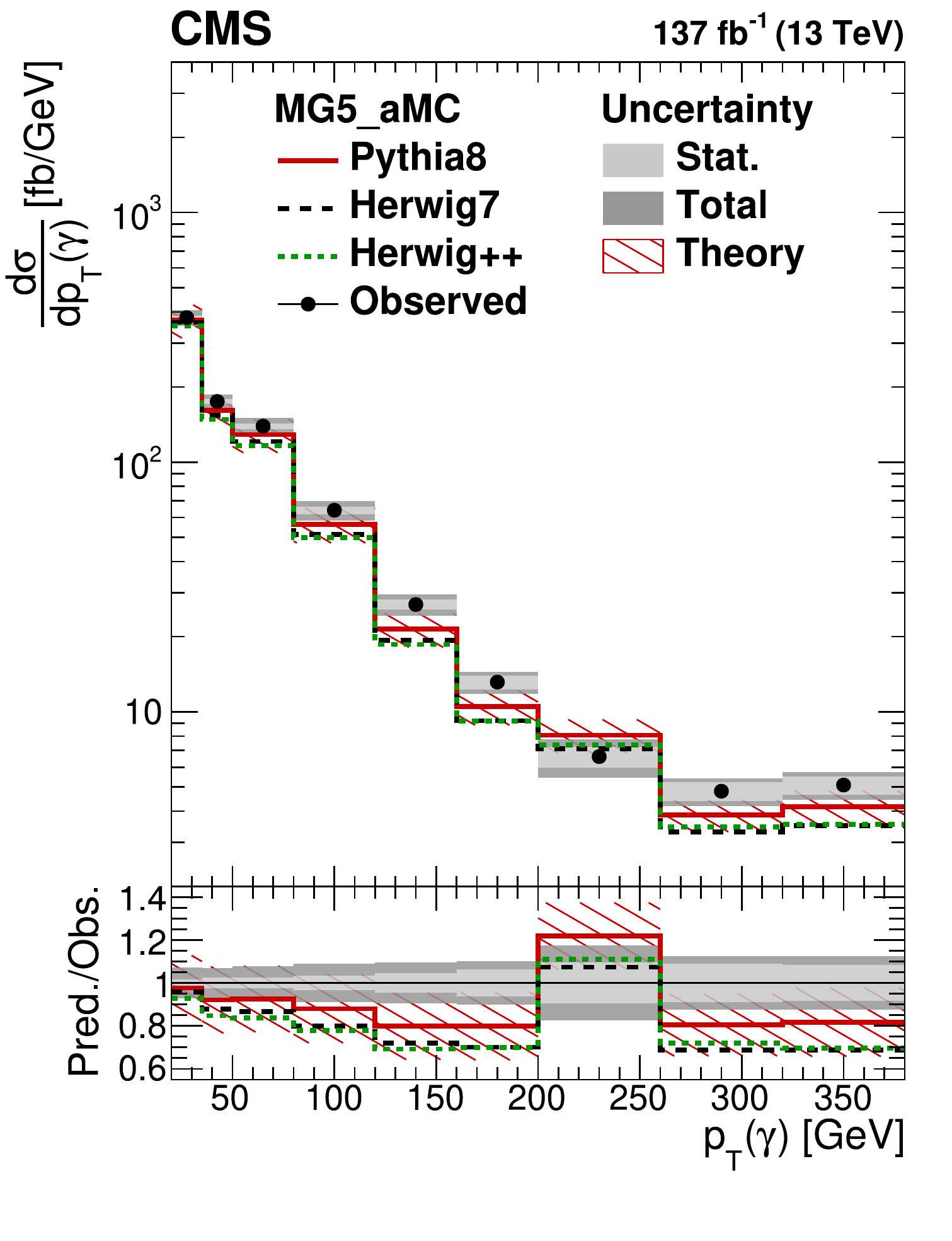}
    \caption{Left: Events in the signal region binned in the distribution that is used for the extraction of the inclusive cross section. Right: Unfolded differential distribution of the photon \pt\ at particle level compared to theory predictions with different shower models ~\cite{ttg_CMS}.}
\label{fig:ttg_CMS}
\end{figure}

The CMS analysis measures the \ttg\ production cross section in a fiducial phase space defined at particle level, where the photon is required not to originate from hadron decays~\cite{ttg_CMS}.
A signal region is defined for events with a single lepton (e or $\mu$), a photon, and at least three jets of which at least one is b tagged. 
Control samples in data are used to estimate backgrounds contributing via nonprompt leptons or photons. 
For processes with a real photon (W$\gamma$ or Z$\gamma$) or where the photon originates from the conversion of an electron, Monte Carlo simulation is used.
Dedicated control regions are defined and included in the signal extraction to constrain the normalization of these contributions.
A maximum likelihood fit is performed, binned for the signal region as shown in Fig.~\ref{fig:ttg_CMS} in the lepton flavor category, jet multiplicity, and the invariant mass of the three-jet system that maximizes the vectorial summed transverse momentum. 
The inclusive cross section is measured as $\sigma_{\ttg} = 800\text{ } \pm 7\text{ (stat) } \pm 46 \text{ (syst) } \mathrm{fb}$, in agreement with the SM prediction of $\sigma_{\ttg}^{\mathrm{SM}} = 770\text{ } \pm 140 \,\mathrm{fb}$ at next-to-leading order (NLO) in quantum chromodynamics (QCD).
Differential cross section measurements at particle level are performed for the transverse momentum \pt\ (shown in Fig.~\ref{fig:ttg_CMS}), and absolute pseudorapidity $|\eta|$ of the photon as well as for the $\delR=\sqrt{\left(\Delta \eta\right)^2 + \left(\Delta \phi\right)^2}$, where $\Delta \eta$ and $\Delta \phi$ are the differences in $\eta$ and the azimuthal angle $\phi$ between the lepton and the photon.
Results are compared with predictions from different shower models and show a preference for the simulation with Pythia 8. 

\begin{figure}[!hbtp]
\centering
    \includegraphics[width=0.45\textwidth]{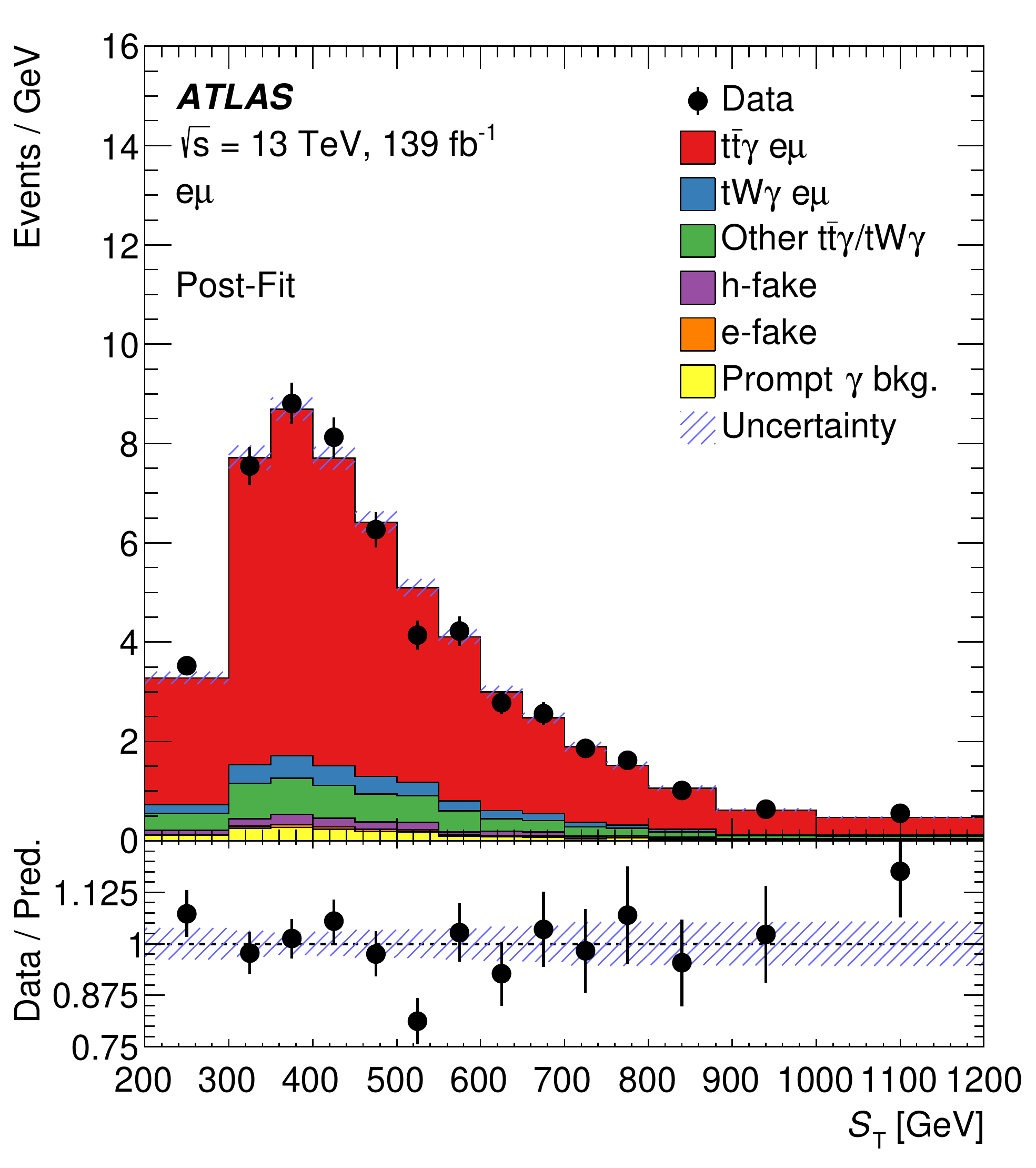}
    \includegraphics[width=0.52\textwidth]{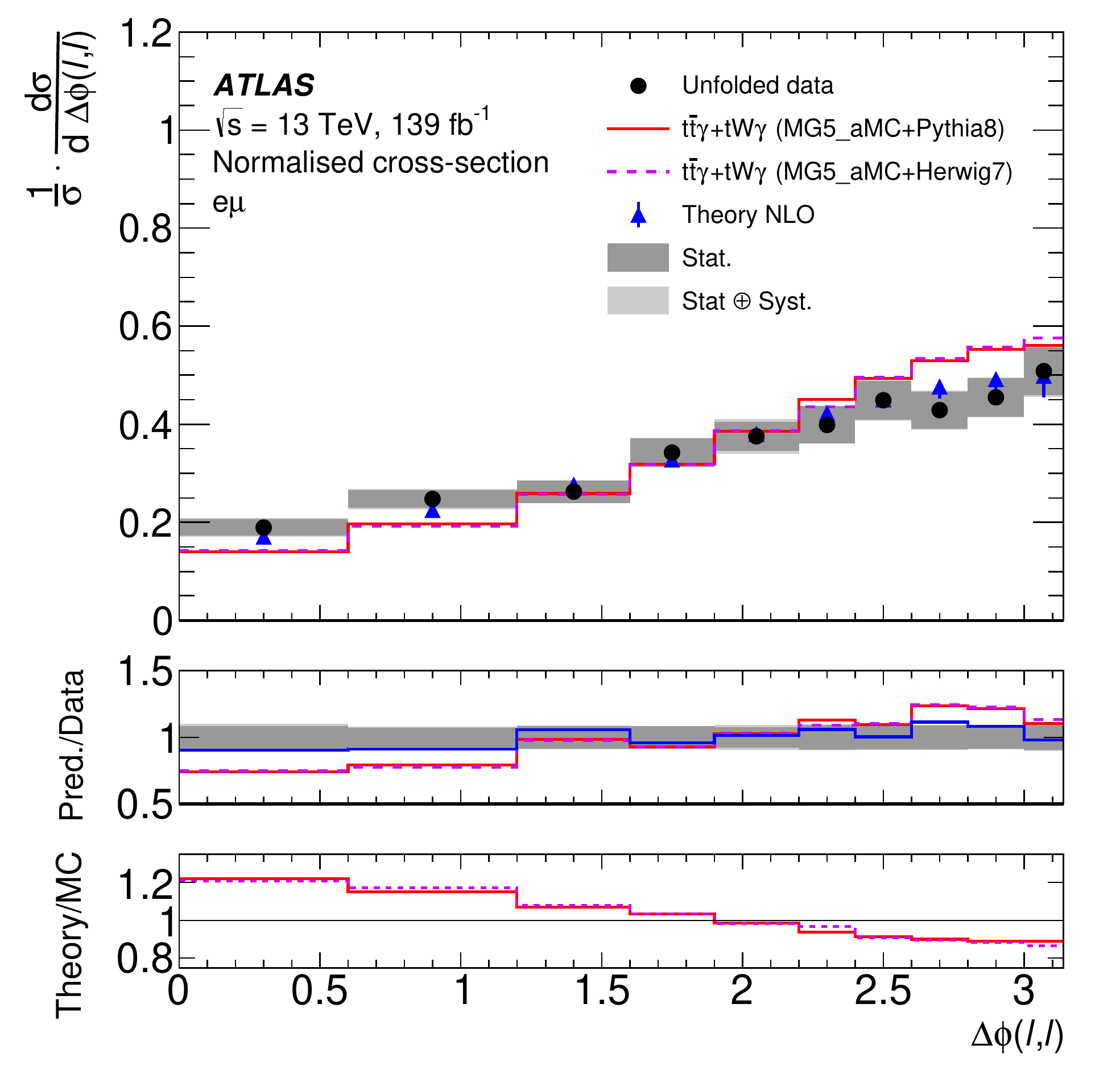}
    \caption{Left: Events in the signal region binned in the distribution that is used for the extraction of the inclusive cross section. Right: Unfolded differential distribution of the $\Delta\phi$ between the leptons in the event at parton level compared to LO and NLO theory predictions~\cite{ttg_ATLAS}.}
\label{fig:ttg_ATLAS}
\end{figure}

The ATLAS measurement is performed in the dilepton final state with exactly one muon and one electron~\cite{ttg_ATLAS}. 
In contrast to the CMS result, the signal definition includes also \twg\ processes and can be compared to the dedicated theory prediction of $\sigma_\mathrm{fid}^{\mathrm{SM}} = 38.5\text{ }^{+0.56}_{-2.18}\text{ (scale) }^{+1.04}_{-1.18}\text{ (PDF) fb}$~\cite{ttg_Theory}, where scale corresponds to the factorization and renormalization scale uncertainty in the matrix element, and PDF stands for the uncertainty from the parton distribution functions.
A binned maximum likelihood fit is performed on the $S_\mathrm{T}$ distribution, which is defined as the scalar sum of the \pt\ of all leptons, photons, jets, and the missing transverse energy in the event. 
The extracted cross section in the fiducial phase space of $\sigma_\mathrm{fid} = 39.6\text{ }\pm 0.8\text{ (stat) } ^{+2.6}_{-2.2}\text{ (syst) } \mathrm{fb}$ is in agreement with the prediction. 
Limiting uncertainties are associated to the modeling of parton shower and initial state radiation of the \ttg\ signal.
The cross section is measured differentially at parton level as a function of the \pt\ and $|\eta|$ of the photon, the $\Delta\phi$ (See Fig.~\ref{fig:ttg_ATLAS}) and $\Delta\eta$ of the two leptons and the minimum \delR\ between any lepton of the event and the photon. 
Comparisons to SM predictions with leading order (LO) and NLO precision in QCD are performed where only the latter one shows good agreement with the data.

\section{FCNC in single top quark production in association with a photon}

A FCNC search is presented by the ATLAS collaboration using $81\,\mathrm{fb}^{-1}$ of data based on the production of a single top quark in association with a photon~\cite{tg_ATLAS}. 
Events with one electron or muon, one photon, one jet that has to be b tagged, and missing transverse energy are selected. 
Contributions where electrons or hadrons are misidentified as photons are estimated from control samples in data.
A neural network (NN) is trained to discriminate SM processes from the FCNC production ($\mathrm{u/c} \rightarrow \mathrm{t}\gamma$) and a fit for the SM hypothesis is performed on the NN score. 
Since no excess in data is observed, limits on the effective couplings are imposed in the EFT framework, the production ($\mathrm{u/c} \rightarrow \mathrm{t}\gamma$), and decay ($\mathrm{t} \rightarrow \mathrm{u/c}\gamma$) for left handed and right handed couplings.

\section{Top quark pair production in association with a Z boson}
The \ttz\ production has been measured in CMS using $77 \, \mathrm{fb}^{-1}$~\cite{ttZ_CMS} and in ATLAS using the full Run-2 dataset of $139\,\mathrm{fb}^{-1}$~\cite{ttZ_ATLAS}. 
Both analyses focus on final states with three or four prompt leptons (e or $\mu$). 
Background processes that contribute with nonprompt leptons are estimated using control samples in data.
Background processes that contribute via prompt leptons, such as diboson production (WZ, ZZ), are estimated from simulation and their uncertainties are constrained by including dedicated control regions in the signal extraction. 
The signal extraction is done via maximum likelihood fits, binned in the number of leptons, the number of jets and b-jets, and, in the ATLAS analysis, also considering the lepton flavors for events with four leptons. 
The obtained inclusive cross sections are $\sigma_{\mathrm{t\bar{t}Z}} = 0.95\text{ } \pm 0.05 \text{ (stat) } \pm 0.06 \text{ (syst) pb}$ and $\sigma_{\mathrm{t\bar{t}Z}} = 0.99\text{ } \pm 0.05 \text{ (stat) } \pm 0.08 \text{ (syst) pb}$ for the CMS and ATLAS measurement, respectively, in agreement with the dedicated theory prediction of $\sigma_{\mathrm{t\bar{t}Z}}^\mathrm{SM} = 0.86\text{ }^{+0.07}_{-0.08} \text{ (scale) } \pm 0.03 \text{ (PDF) pb}$
including NLO in QCD and EW, and NNLL terms~\cite{ttZ_Theory}. 
Leading systematic uncertainties are related to the parton shower modeling, modeling of background processes, lepton identification (CMS), and b-jet identification (ATLAS).

\begin{figure}[!hbtp]
\centering
    \includegraphics[width=0.49\textwidth]{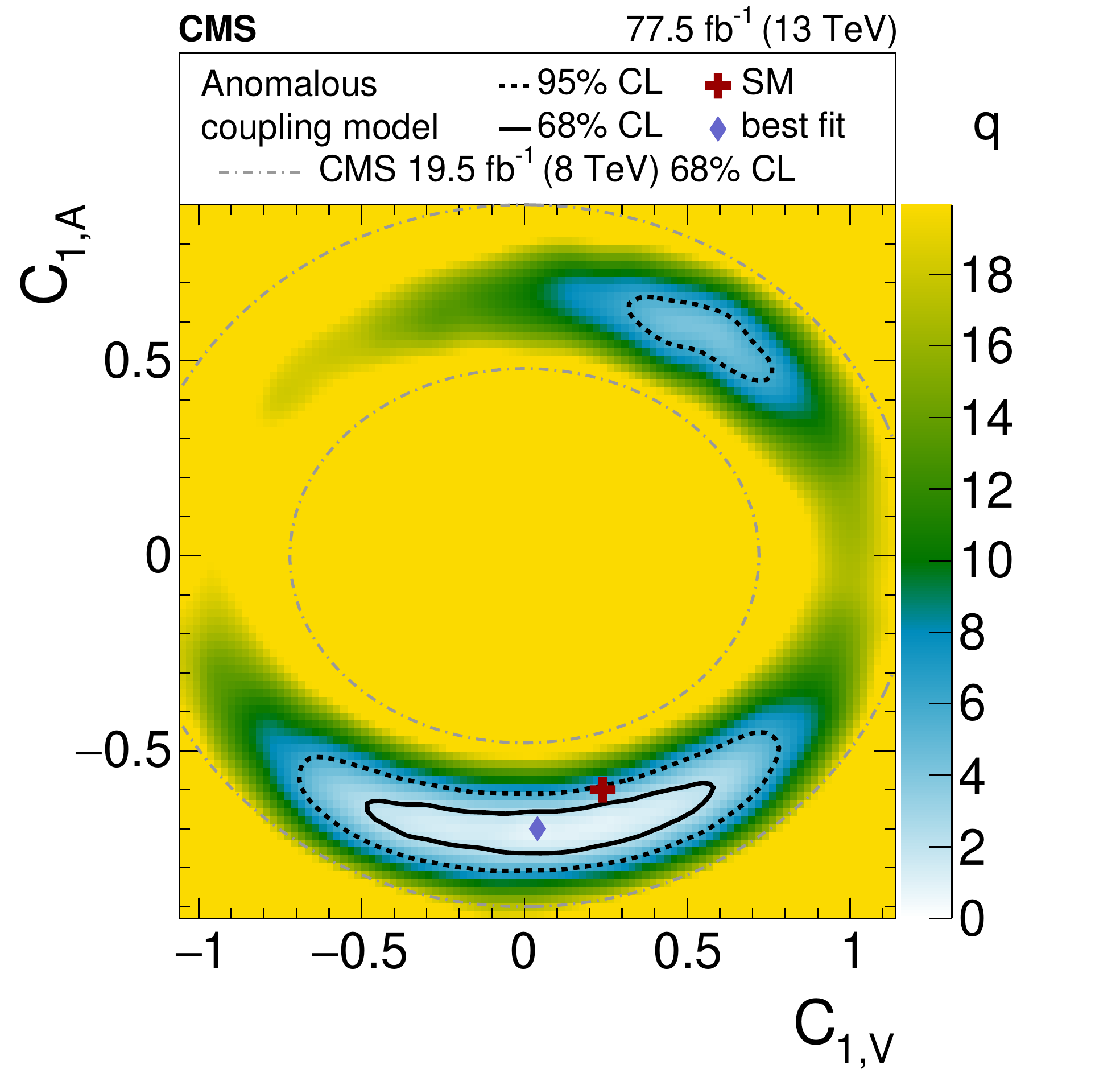}
    \includegraphics[width=0.5\textwidth]{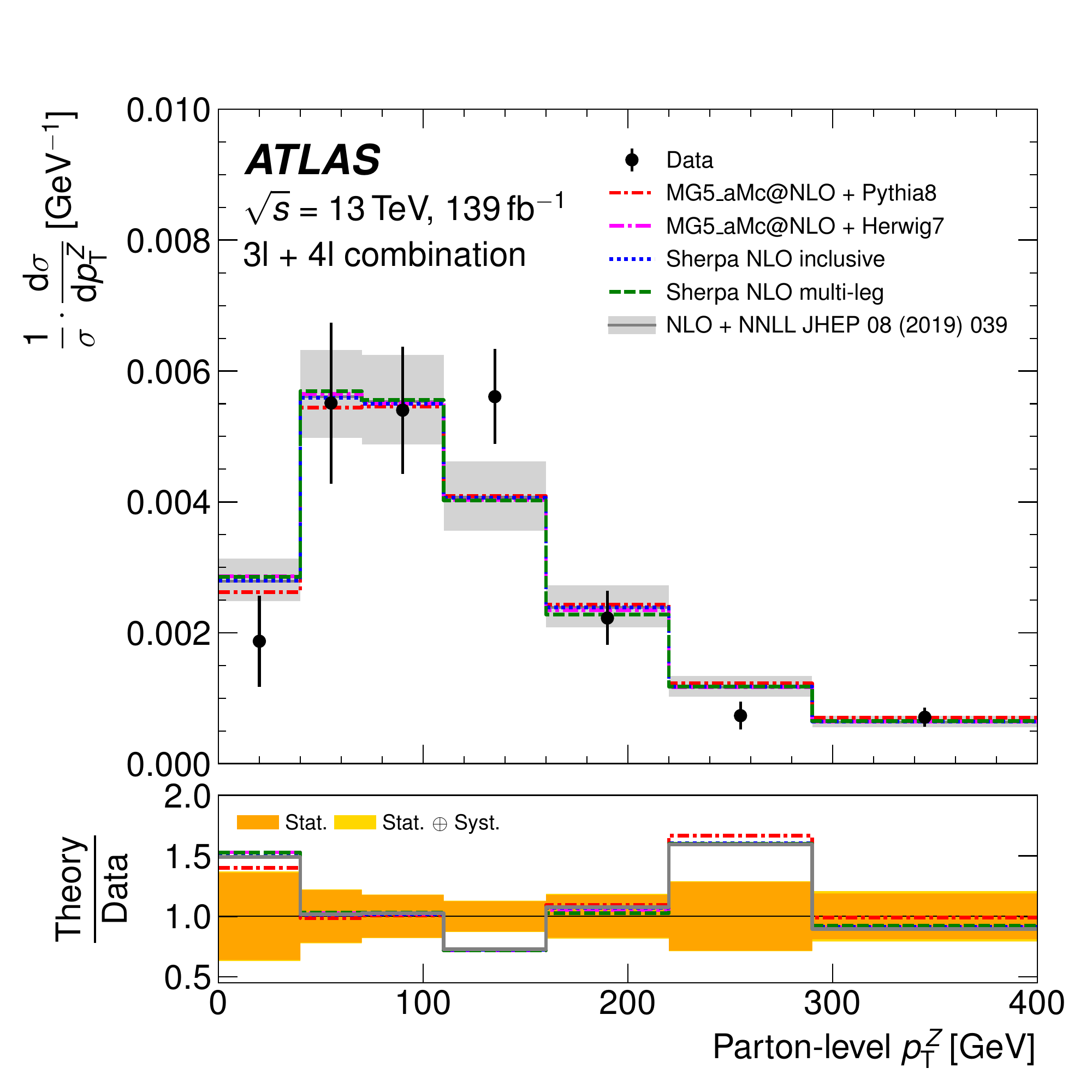}
    \caption{Left: A two-dimensional scan of the log likelihood ratio q with respect to the best-fit value for axial-vector $\mathrm{C}_{1,\mathrm{A}}$ and vector current $\mathrm{C}_{1,\mathrm{V}}$ couplings from the \ttz analysis of CMS~\cite{ttZ_CMS}.
    Right: The differential distribution of the Z boson \pt\ at parton level from the \ttz\ analysis of ATLAS~\cite{ttZ_ATLAS}.}
\label{fig:ttz}
\end{figure}

In the CMS analysis, limits on anomalous neutral current interactions and EW dipole moments are extracted as shown in Fig.~\ref{fig:ttz}. 
Simulated samples are reweighted and the fit is repeated with a binning, chosen to be more sensitive to these effects. 
In particular, events are binned in the \pt\ of the Z boson and in the cosine of the angle between the negatively charged lepton and the Z boson in the Z boson rest frame.
These two variables are also chosen for a first differential cross section measurements of \ttz. 
Including events with three leptons and at least three jets of which at least one has to be b tagged, the distributions are unfolded to parton level using matrix inversion techniques.
Comparisons with theory calculations show good agreement.

The ATLAS analysis includes differential cross section measurements at both, parton and particle level, absolute and normalized. 
Various observables are chosen, sensitive to generator modeling, QCD effects, top \pt\ modeling, BSM effects, top-Z vertex, or spin correlation.
Distributions are unfolded using an iterative Bayesian approach for either events with three or four leptons separately, or all together. 
The unfolded distribution of the Z boson \pt\ is shown in Fig.~\ref{fig:ttz}.
For each distribution, the systematic uncertainties are studied and results are compared to simulations using different generators and parton shower models. 
Overall a good agreement is observed.

\section{Single top quark production in association with a Z boson}
\begin{figure}[!hbtp]
\centering
    \includegraphics[width=0.455\textwidth]{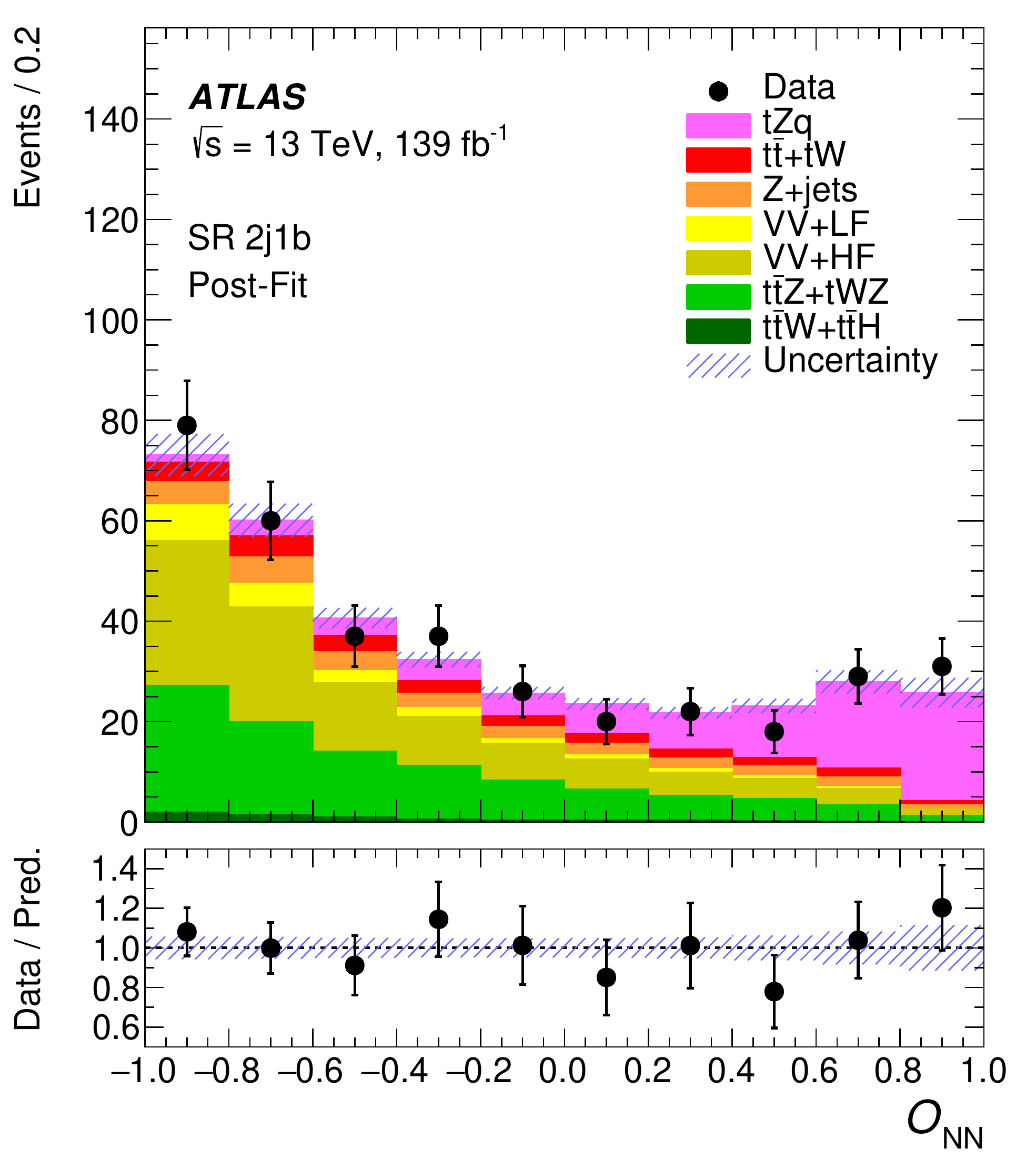}
    \includegraphics[width=0.535\textwidth]{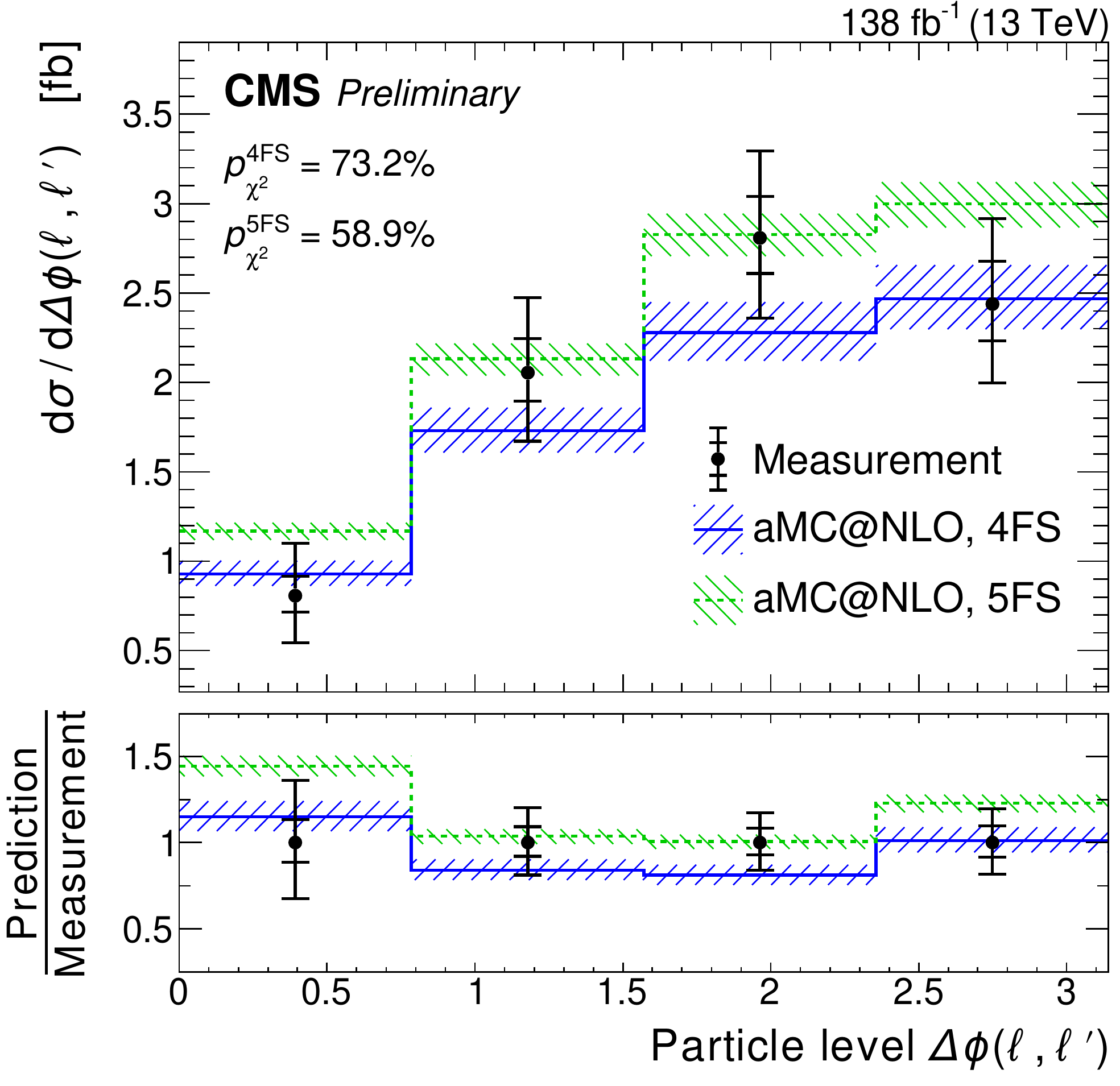}
    \caption{Left: Discriminator distribution of the NN output in the signal region with two jets for the inclusive cross section measurement by ATLAS~\cite{tZq_ATLAS}.
    Right: Unfolded differential cross section as a function of the angular azimuthal difference between the two leptons from the Z boson at particle level from the CMS measurement~\cite{tZq_CMS}.}
\label{fig:tzq}
\end{figure}

Measurements of \tzq\ are performed by both ATLAS and CMS experiments, using the full Run-2 data~\cite{tZq_ATLAS,tZq_CMS}.
Events with three leptons (e or $\mu$) and at least two jets with at least one b jet are selected. 
Backgrounds with nonprompt leptons are estimated from control samples in data.
Various control regions are included in the signal extraction to constrain backgrounds with both prompt and nonprompt leptons.
Distinctive features like the charge asymmetry or the presence of a high-\pt\ jet in the forward region of the detector are used to separate the signal from the backgrounds via MVA techniques.
An NN is used in the ATLAS analysis (Shown in Fig.~\ref{fig:tzq}) while CMS employs a BDT for the inclusive cross section measurement.
A fit on the discriminator distributions results in $\sigma_{\mathrm{t{\ell^+}{\ell^{-}}q}} = 97\text{ } \pm 13 \text{ (stat) } \pm 7 \text{ (syst) } \mathrm{fb}$ and $\sigma_{\mathrm{t{\ell^+}{\ell^{-}}q}} = 87.9\text{ }^{+7.5}_{-7.3} \text{ (stat) }^{+7.3}_{-6.0} \text{ (syst) } \mathrm{fb}$ with $m_{{\ell^+}{\ell^{-}}} > 30\,\text{GeV}$ for the ATLAS and CMS measurement, respectively.
Both measurements are in agreement with SM predictions. 

The CMS analysis also includes the first differential cross section measurements of \tzq. 
Therefore an NN is used to isolate the signal from the backgrounds. 
Unfolded distributions are extracted both at parton and particle level via a multidimensional binned maximum likelihood fit. 
Nine different observables are chosen based on their sensitivity to BSM effects and generator modeling. 
Comparisons with SM predictions show good agreement, an example of an unfolded distribution is shown in Fig.~\ref{fig:tzq}.
The top quark spin asymmetry, defined as the asymmetry of an angular distribution between the light spectator quark and the lepton from the top quark, is also measured for the first time in \tzq. 
A value of $0.58\text{ }^{+0.15}_{-0.16}\text{ (stat) } \pm 0.06 \text{ (syst)}$ is measured and found to be in agreement with the SM prediction of $0.437\text{ }^{+0.004}_{-0.003}$.


\section{Summary}
Proton-proton collision data at a center-of-mass energy of 13\,TeV corresponding to an integrated luminosity of up to 138\,$\mathrm{fb}^{-1}$ have been collected in the years between 2015 and 2018 at the CERN LHC. 
These data facilitate for the first time a detailed study of top quark processes in association with vector bosons.
Most resent measurements of top quark pair production in association with a Z boson or a photon (\ttz\ and \ttg), as well as single top quark production in association with a Z boson or a photon (\tzq\ and \tqg) from the ATLAS and CMS experiments are summarized. 
Inclusive cross section measurements are presented with an unprecedented precision that is already systematically limited and competing with the uncertainty of the theory predictions in the case of \ttz\ and \ttg. 
For the first time, differential cross section measurements are performed for the \ttz, \ttg, and \tzq\ processes, including comparisons with different generators using different shower models. 
In general, both the inclusive and differential cross section measurements show good agreement with standard model predictions. 
From the \ttz\ measurement by CMS, also limits on anomalous couplings are obtained while a dedicated search on \tqg\ is done by ATLAS to extract limits on FCNC coupling and parameters in the EFT framework.

For some processes, the full available LHC Run-2 dataset has not yet been used and various decay channels are not analyzed yet.
The data also allow for the combined measurements of different processes and the combination of results from CMS and ATLAS. 
Therefore, we can expect much more complementary results and more precise measurements by both ATLAS and CMS experiments in the near future.

\newcommand{\arxiv}[1]{\href{http://arxiv.org/abs/#1}{#1}}
\newcommand{\cds}[1]{\href{http://cds.cern.ch/record/#1}{CDS:#1}}
\newcommand{\doi}[2]{\href{http://doi.org/#1}{\mbox{#2}}}
\newcommand{\pas}[1]{\href{http://cms-results.web.cern.ch/cms-results/public-results/preliminary-results/#1/index.html}{CMS-PAS-#1}}


\begin{thebibliography}{99}
\small
\setlength\parskip{-1.2ex}

\bibitem{ATLAS}
    ATLAS Collaboration. 
    \emph{The {ATLAS} Experiment at the {CERN} Large Hadron Collider},
    \doi{10.1088/1748-0221/3/08/s08003}{\emph{JINST} \textbf{03} (2008) S08003}

\bibitem{CMS}
    CMS Collaboration.
    \emph{The {CMS} experiment at the {CERN} {LHC}},
    \doi{10.1088/1748-0221/3/08/s08004}{\emph{JINST} \textbf{03} (2008) S08004}

\bibitem{ttg_CMS}
    CMS Collaboration.
    \emph{Measurement of the inclusive and differential $\mathrm{t\overline{t}}\gamma$ cross sections in the single-lepton channel and {EFT} interpretation at $\sqrt{s}$ = 13\,{TeV}}, 
    \texttt{\arxiv{2107.01508}}
    (submitted to \emph{JHEP}).

\bibitem{ttg_ATLAS}
    ATLAS Collaboration.
    \emph{Measurements of inclusive and differential cross-sections of combined $ \mathrm{t}\overline{\mathrm{t}}\gamma$ and $\mathrm{tW}\gamma$ production in the e$\mu$ channel at 13\,{TeV} with the {ATLAS} detector},
    \doi{10.1007/jhep09(2020)049}{\emph{JHEP} \textbf{09} (2020) 049} 
    [\texttt{\arxiv{2007.06946}}].

\bibitem{ttg_Theory}
    G.~Bevilacqua {\it et al}.
    \emph{Hard photons in hadroproduction of top quarks with realistic final states},
    \doi{10.1007/jhep10(2018)158}{\emph{JHEP} \textbf{10} (2018) 158}
    [\texttt{\arxiv{1803.09916}}].

\bibitem{tg_ATLAS}
    ATLAS Collaboration.
    \emph{Search for flavour-changing neutral currents in processes with one top quark and a photon using 81\,$\mathrm{fb}^{-1}$ of pp collisions at $\sqrt{s}=13$\,{TeV} with the {ATLAS} experiment},
    \doi{10.1016/j.physletb.2019.135082}{\emph{Phys.Lett.B} \textbf{800} (2020) 135082}
    [\texttt{\arxiv{1908.08461}}].

\bibitem{ttZ_CMS}
    CMS Collaboration.
    \emph{Measurement of top quark pair production in association with a {Z} boson in proton-proton collisions at $\sqrt{s} = 13$\,{TeV}},
    \doi{10.1007/jhep03(2020)056}{\emph{JHEP} \textbf{03} (2020) 056}
    [\texttt{\arxiv{1907.11270}}].

\bibitem{ttZ_ATLAS}
    ATLAS Collaboration.
    \emph{Measurements of the inclusive and differential production cross sections of a top-quark-antiquark pair in association with a {Z} boson at $\sqrt{s} = 13$\,{TeV} with the {ATLAS} detector}, 
    \doi{10.1140/epjc/s10052-021-09439-4}{\emph{Eur. Phys. J. C} \textbf{81} (2021) 737}
    [\texttt{\arxiv{2103.12603}}].


\bibitem{ttZ_Theory}
    A.~Kulesza {\it et al}.
    \emph{Associated top quark pair production with a heavy boson: differential cross sections at {NLO} + {NNLL} accuracy},
    \doi{10.1140/epjc/s10052-020-7987-6}{\emph{Eur. Phys. J. C} \textbf{80} (2020) 428}
    [\texttt{\arxiv{2001.03031}}].

\bibitem{tZq_ATLAS}
    ATLAS Collaboration.
    \emph{Observation of the associated production of a top quark and a {Z} boson in pp collisions at $\sqrt{s} = 13$\,{TeV} with the {ATLAS} detector},
    \doi{10.1007/jhep07(2020)124}{\emph{JHEP} \textbf{07} (2020) 124}
    [\texttt{\arxiv{2002.07546}}].

\bibitem{tZq_CMS}
    CMS Collaboration.
    \emph{Inclusive and differential cross section measurements of single top quark production in association with a Z boson in proton-proton collisions at $\sqrt{s} = 13$\,{TeV}},
    \texttt{\pas{TOP-20-010}},
    CERN 2020
    [\texttt{\cds{2771809}}]

\end{thebibliography}
\end{document}